\documentclass[aps,twocolumn,nofootinbib,floatfix]{revtex4}
\usepackage{amsfonts}
\usepackage{amssymb}
\usepackage{amsmath}
\usepackage{graphics}
\usepackage{graphicx}
\usepackage{soul}
\usepackage{natbib}
\usepackage{bm}
\usepackage{mathtools}
\usepackage{epstopdf}
\usepackage{color}
\usepackage{lipsum}
\usepackage[colorlinks=true,citecolor=black,linkcolor=black,urlcolor=black,filecolor=black]{hyperref}

\usepackage{acronym}
\newacro{DF}{distribution function}
\newacroplural{DF}[DFs]{distribution functions}
\newcommand{\DF}{\ac{DF}}

\newacro{BL}{Balescu--Lenard}
\newcommand{\BL}{\ac{BL}}
\newacro{HMF}{Hamiltonian Mean Field}
\newcommand{\HMF}{\ac{HMF}}

\newcommand{\rd}{\mathrm{d}}
\newcommand{\ri}{\mathrm{i}}
\newcommand{\re}{\mathrm{e}}
\newcommand{\bL}{\mathbf{L}}
\newcommand{\Uext}{U_{\mathrm{ext}}}
\newcommand{\Mtot}{M_{\mathrm{tot}}}
\newcommand{\ellp}{\ell^{\prime}}
\newcommand{\mpp}{m^{\prime}}
\newcommand{\bX}{\mathbf{X}}
\newcommand{\bXext}{\mathbf{X}_{\rm ext}}
\newcommand{\deltaD}{\delta_{\mathrm{D}}}
\newcommand{\psidtot}{\psi^{\mathrm{d}}_{\mathrm{tot}}}
\newcommand{\up}{u^{\prime}}
\newcommand{\psid}{\psi^{\mathrm{d}}}
\newcommand{\wbM}{\widehat{\mathbf{M}}}

\newcommand{\kp}{k^{\prime}}
\newcommand{\alephD}{\aleph_{\mathrm{D}}}
\newcommand{\bw}{\mathbf{w}}
\newcommand{\bwp}{\mathbf{w}^{\prime}}
\newcommand{\bT}{\bm{\theta}}
\newcommand{\bTp}{\bm{\theta}^{\prime}}
\newcommand{\bJ}{\mathbf{J}}
\newcommand{\bJp}{\mathbf{J}^{\prime}}
\newcommand{\bO}{\mathbf{\Omega}}
\newcommand{\bLp}{\mathbf{L}^{\prime}}
\newcommand{\vtheta}{\vartheta}
\newcommand{\bk}{\mathbf{k}}
\newcommand{\bkp}{\mathbf{k}^{\prime}}
\newcommand{\bI}{\mathbf{I}}

\newcommand{\Dext}{D_{\mathrm{ext}}}
\newcommand{\veps}{\varepsilon}
\newcommand{\mO}{\mathcal{O}}
\newcommand{\ot}{\overline{t}}
\newcommand{\om}{\overline{m}}
\newcommand{\tw}{\widetilde{\omega}}

\begin{document}

\title{Secular dynamics of long-range interacting particles on a sphere
\\
in the axisymmetric limit}

\author{Jean-Baptiste Fouvry\footnote{Hubble Fellow}}
\affiliation{Institute for Advanced Study, Princeton, NJ, 08540, USA}
\author{Ben Bar-Or}
\affiliation{Institute for Advanced Study, Princeton, NJ, 08540, USA}
\author{Pierre-Henri Chavanis}
\affiliation{Laboratoire de Physique Th\'eorique, Universit\'e de Toulouse, CNRS, UPS, France}

\begin{abstract}
  We investigate the secular dynamics of long-range interacting particles
  moving on a sphere, in the limit of an axisymmetric mean field potential.
  We show that
  this system can be described by the general kinetic equation, the
  inhomogeneous Balescu--Lenard equation. We use this approach to
  compute long-term diffusion coefficients, that are compared with direct
  simulations. Finally, we show how the scaling of the system's relaxation rate
  with the number of particles fundamentally depends on the underlying
  frequency profile. This clarifies why systems with a monotonic profile
  undergo a kinetic blocking and cannot relax as a whole under ${1/N}$ resonant
  effects.  Because of its general form, this framework can describe the
  dynamics of globally coupled classical Heisenberg spins, long-range couplings
  in liquid crystals, or the orbital inclination evolution of stars in nearly
  Keplerian systems.
\end{abstract}
\maketitle

\section{Introduction}
\label{sec:Introduction}

Long-range interacting systems generically undergo an evolution in two
stages. First, a fast (collisionless) violent relaxation~\citep{LyndenBell1967}
during which the system reaches a quasistationary state (a steady
state of the collisionless Boltzmann equation) and the system is dynamically
frozen under the mean field dynamics. Then, as a consequence of the finite
number of particles, the system undergoes a slow (collisional) relaxation that
drives it towards thermodynamical equilibrium.  This second stage is
generically described by the \BL\ equation~\citep{Balescu1960,Lenard1960},
recently generalized to inhomogeneous
systems~\citep{Heyvaerts2010,Chavanis2012}. These formalisms can account
simultaneously for inhomogeneity (i.e.\ non-trivial orbital structures),
collective effects (i.e., spontaneous amplification of perturbations) and
non-local resonant couplings.

In this letter, we focus our attention on one such long-range interacting
system, namely the problem of long-range coupled particles evolving on a
sphere. Because of its general form, this system is of relevance in various
physical setups ranging from spin dynamics to stellar systems (see
Section~\ref{sec:Model}). Here, we show how in the axisymmetric limit, the
generic methods of inhomogeneous kinetic theory can be applied, and accordingly
derive the associated kinetic equation. In addition to allowing for
quantitative predictions of the system's diffusion coefficients, we clarify how
this theory predicts the dependence of the system's relaxation rate with the
number of particles, and the important role played by the frequency profile in
that respect.

The paper is organized as follows. In Section~\ref{sec:Model}, we present the
considered model. Placing ourselves within the axisymmetric limit, we derive in
Section~\ref{sec:BLEq} the appropriate Balescu--Lenard equation describing the
long-term evolution of that system. In Section~\ref{sec:Application}, we
present applications of this formalism to recover the system's diffusion
coefficients as well the scaling of the relaxation rate with the number of
particles. Finally, we conclude in Section~\ref{sec:Conclusion}.

\section{The model}
\label{sec:Model}

We consider a set of $N$ particles evolving on a sphere of unit radius, and denote the
spherical coordinates with ${ (\phi , \vtheta) }$. To any location on the
sphere, we associate a normal vector ${ \bL = \bL (\phi , \vtheta) }$. The
specific Hamiltonian of the system is
\begin{equation}
  \label{def_Hamiltonian}
  H = \mu \sum_{i < j} U (\bL_{i} \cdot \bL_{j}) + \sum_{i} \Uext (\bL_{i}),
\end{equation}
where ${ \mu = \Mtot / N }$ is the individual mass of the particles,
${ U (\bL, \bL') = U (\bL \cdot \bL') }$ is the
pairwise interaction, and ${ \Uext(\bL) }$ is an imposed external potential.  The pairwise
interaction is developed in Legendre Polynomials as
\begin{align}
  \label{Legendre_U}
  U (\bL_{i} \cdot \bL_{j}) 
  & 
    \, = - \sum_{\ell} \alpha_{\ell} P_{\ell} (\bL_{i} \cdot \bL_{j})
  \\ \nonumber
  & 
    \, = - \sum_{\ell , m} \alpha_{\ell} \, b_{\ell} \, Y_{\ell}^{m} (\bL_{i}) \, Y_{\ell}^{m *} (\bL_{j}) ; \;\; b_{\ell} = \frac{4 \pi}{2 \ell + 1} ,
\end{align}
where we used the addition theorem, and introduced the spherical harmonics
${ Y_{\ell}^{m} (\bL) = K_{\ell}^{m} P_{\ell}^{m} (u) \, \re^{\ri m \phi} }$,
where ${ P_{\ell}^{m} (u) }$ is the associated Legendre
functions~\citep{Arfken2013}, and
${ K_{\ell}^{m} \!=\! \big[ \tfrac{2 \ell + 1}{4 \pi} \tfrac{(\ell -
      m)!}{(\ell + m)!} \big]^{1/2} }$.  The spherical harmonics are
normalized as
${ \!\int\! \rd \bL Y_{\ell}^{m} Y_{\ellp}^{\mpp *} \!=\! \delta_{\ell}^{\ellp}
  \delta_{m}^{\mpp} }$, with the unit volume
${ \rd \bL = \rd \vtheta \sin (\vtheta) \rd \phi }$.  The canonical coordinates
associated with this two-dimensional phase space are
${ \bw = (\phi , \cos (\vtheta) = u) }$, and the equations of motion for
particle $i$ are ${ \dot{\phi}_{i} = \partial H/\partial u_{i} }$ and
${ \dot{u}_{i} = - \partial H / \partial \phi_{i} }$. We recast these equations
as
\begin{equation}
\label{EOM_short}
\frac{\rd \bL_{i}}{\rd t} = \sum_{\ell , m} M_{\ell}^{m} (t) \, \bX_{\ell}^{m} (\bL_{i}) + \bXext (\bL_{i}) ,
\end{equation}
where
${ \bX_{\ell}^{m} (\bL) = \bL \times \partial Y_{\ell}^{m} / \partial \bL }$
are the vector spherical harmonics, 
\begin{equation}
M_{\ell}^{m} (t) = \mu \, \alpha_{\ell} \, b_{\ell} \sum_{j} Y_{\ell}^{m *} (\bL_{j} (t)) 
\label{def_Mlm}
\end{equation}
are the system's
instantaneous magnetisations
and
${ \bXext (\bL) = - \bL \times \partial \Uext / \partial \bL }$ captures the
contribution from the external potential. 

Equation~\eqref{EOM_short} is the exact evolution equation of this problem.
The Hamiltonian from Eq.~\eqref{def_Hamiltonian} encompasses a wide class of
long-range interacting systems: (i) ${ U = - \alpha_{1} P_{1} }$ describes
globally coupled classical Heisenberg
spins~\citep{GuptaMukamel2011,BarreGupta2014}, (ii)
${ U = - \alpha_{2} P_{2} }$ is the Maier-Saupe model for liquid
crystals~\citep{MaierSaupe1958,RoupasKocsis2017}, (iii)
${ U = - \sum_{\ell} \alpha_{2\ell} P_{2\ell} }$ captures the
process of \textit{vector resonant relaxation} in galactic
nuclei~\citep{KocsisTremaine2015,SzolgyenKocsis2018,TakacsKocsis2018} (up to
additional conserved quantities).

In the coming section, we consider the most general setup, but place ourselves
within the axisymmetric limit, i.e.\ where the mean field Hamiltonian is
invariant w.r.t.\ $\phi$.  We show how the general kinetic theory of long-range
interacting systems (see Appendix~\ref{sec:GenericBL}) can straightforwardly be
applied to this regime, and accordingly derive the associated evolution
equation.\footnote{We make the correspondence with
  Appendix~\ref{sec:GenericBL} by noting that ${ \bL = (\phi , u) }$ plays the
  role of ${ \bw = (\theta , J) }$, $\phi$ the role of the angle $\theta$, and
  $u$ the role of the action $J$.  }

\section{The Balescu--Lenard equation}
\label{sec:BLEq}

Let us assume that the system is characterized by a mean \DF\@, ${ F (\bL) }$,
normalized so that ${ \!\int\! \rd \bL F = M_{\rm tot} }$, with
${ M_{\rm tot} = 1 }$ the total mass of the system. Following
Eq.~\eqref{def_Hamiltonian}, the mean specific Hamiltonian of a particle in
that system reads
\begin{align}
H_{0} (\bL) & \, = \!\! \int \!\! \rd \bLp \, U (\bL \cdot \bLp) \, F (\bLp) + \Uext (\bL) 
\nonumber
\\
& \, = \sum_{\ell} h_{\ell} P_{\ell} (u) + \Uext (u),
\label{mean_Hamiltonian}
\end{align}
where in the second line, we assumed that the system's \DF\ and the external
potential are axisymmetric, i.e., ${ F (\bL) = F (u) }$ and
${ \Uext (\bL) = \Uext (u) }$, and introduced the coefficients
${ h_{\ell} = - 2 \pi \alpha_{\ell}  \!\int\! \rd \up P_{\ell}
  (\up) F(\up) }$. The associated orbital frequency
${ \Omega (u) = \rd H_{0} / \rd u }$ naturally follows from
Eq.~\eqref{mean_Hamiltonian}. For axisymmetric configurations, we have
${ H_{0} (\bL) = H_{0} (u) }$. Therefore, the Poisson bracket satisfies
${ [ H_{0} (u) , F (u) ] = 0 }$, i.e., any axisymmetric \DF\ is a steady state
for the mean field dynamics. In addition, the mean Hamiltonian is integrable,
as the action ${ J = u }$ is conserved along the mean motion, while the
associated angle ${\theta = \phi }$, evolves linearly in time with the
frequency ${ \Omega (u) }$.

Investigating the long-term evolution of such a quasi-stationary steady amounts
to investigating the slow distortion of the system's mean \DF, ${ F (u) }$
(assumed to remain linearly stable and axisymmetric throughout its evolution).
Following the general kinetic theory of long-range interacting integrable
systems (briefly reproduced in Appendix~\ref{sec:GenericBL}), deriving the
kinetic equation for ${ F (u) }$ is immediate.  One only needs to proceed by
analogies as we detail below.

The interaction potential can be written under the separable form
${ U (\bL \cdot \bLp) = - \sum_{p} \psi^{(p)} (\bL) \, \psi^{(p) *} (\bLp) }$, where
the potential basis elements are
\begin{equation}
\psi^{(p)} (\bL) = C_{\ell^{p}} \, Y_{\ell^{p}}^{m^{p}} (\bL) , \;\; C_{\ell} =
\sqrt{\alpha_{\ell} b_{\ell}}.
\label{def_psip}
\end{equation}
Fourier transform w.r.t.\ the angle ${ \theta = \phi }$ reads
\begin{equation}
\psi^{(p)}_{k} (u) = \!\! \int \!\! \frac{\rd \phi}{2 \pi} \, \re^{- \ri k \phi} \, C_{\ell^{p}} \, Y_{\ell^{p}}^{m^{p}} (u , \phi) = \delta_{k}^{m^{p}} c_{\ell^{p}}^{m^{p}} (u) ,
\label{FT_basis}
\end{equation}
with the coefficient ${ c_{\ell}^{m} (u) = C_{\ell} \, K_{\ell}^{m} \, P_{\ell}^{m} (u) }$.
Injected in Eq.~\eqref{Fourier_M_generic}, the system's response matrix becomes
\begin{equation}
\label{Fourier_M}
\wbM_{pq} (\omega) = 2 \pi \, \delta_{m^{p}}^{m^{q}} \!\! \int \!\! \rd u \, \frac{m^{p} \partial F / \partial u}{\omega - m^{p} \Omega (u)} \, c_{\ell^{p}}^{m^{p} *} (u) \, c_{\ell^{q}}^{m^{q}} (u).
\end{equation}
A system is then said to be linearly unstable if there exists a complex
frequency ${ \omega = \omega_{0} + \ri \eta }$ (with ${ \eta > 0 }$), for which
${ \wbM (\omega) }$ admits an eigenvalue equal to $1$.  In that case, the
system supports an unstable mode of pattern speed $\omega_{0}$, and growth rate
$\eta$~\citep[see Section~{5.3} in][]{BinneyTremaine2008}. In present context, Eq.~\eqref{Fourier_M} generalizes the stability criteria
put forward in~\cite{GuptaMukamel2011,BarreGupta2014} (see
Appendix~\ref{sec:HeisenbergLimit}).

Following Eq.~\eqref{dressed_psi}, the system's \textit{dressed}
susceptibility coefficients read
\begin{equation}
  \psid_{k \kp} (u , \up , \omega) = - \delta_{k}^{\kp} \!\! \! \sum_{\ell, \ellp \ge |k|}
\! \! c_{\ell}^{k} (u) \, c_{\ellp}^{k *} (\up) \, {\big[\bI_k - \wbM_k (\omega) \big]}^{-1}_{\ell \ellp} ,
\label{suscp_coeff}
\end{equation}
where
\begin{equation}
{\big[ \bI_{k} \big]}_{\ell \ellp} = \delta_{\ell}^{\ellp}; \;\;\; {\big[
  \wbM_{k} (\omega) \big]}_{\ell \ellp} = \wbM_{[\ell , k] , [\ellp , k]} (\omega) .
\label{definition_I_Lambda}
\end{equation}

Assuming that the system is linearly stable, and that the frequency profile is
non-degenerate (i.e.\ ${\partial \Omega / \partial u = 0}$ only in isolated
points), the long-term evolution of this axisymmetric system is characterized
by the inhomogeneous \BL\ equation (see Eq.~\eqref{generic_BL}), that reads
here
\begin{align}
\frac{\partial F}{\partial t} = 2 \pi^{2} \mu \frac{\partial }{\partial u} & \, \bigg[ \!\! \int \!\! \rd \up \, {\big| \psidtot (u , \up , \Omega (u)) \big|}^{2} \, \deltaD (\Omega(u) - \Omega(\up))
\nonumber
\\
& \, \times \bigg( \frac{\partial }{\partial u} - \frac{\partial }{\partial \up} \bigg) \, F (u) \, F (\up) \bigg] ,
\label{BL_Eq}
\end{align}
where we introduced the total dressed susceptibility coefficients ${ \psidtot (u , \up , \omega) }$ as
\begin{equation}
{\big| \psidtot (u , \up , \omega) \big|}^{2} = 2 \sum_{k \geq 1} k \, {\big| \psid_{k k} (u , \up , k \omega) \big|}^{2} .
\label{definition_psidtot}
\end{equation}
In Eq.~\eqref{BL_Eq}, we emphasize the absence of a sum on resonance vectors,
owing to the Kronecker symbol in Eq.~\eqref{suscp_coeff}.  Collective effects
can be switched off by imposing ${ \wbM_{pq} (\omega) = 0 }$ (i.e.\ replacing
the dressed susceptibility coefficients,
${ \psid_{k \kp} (u , \up , \omega) }$, by their bare analogs,
${ \psi_{k\kp} (u , \up) }$, see Eq.~\eqref{separable_U}), which leads to the
inhomogeneous Landau equation~\citep{Chavanis2013}.  Finally, we recall that
Eq.~\eqref{BL_Eq} can be rewritten as a Fokker--Planck equation
\begin{equation}
\frac{\partial F}{\partial t} = - \frac{\partial }{\partial u} \big[ D_{1} (u) \, F (u) \big] + \frac{1}{2} \frac{\partial^{2}}{\partial u^{2}} \big[ D_{2} (u) \, F (u) \big] ,
\label{FP_Form}
\end{equation}
with the first- and second-order diffusion coefficients
\begin{align}
D_{2} (u) & \!=\! {(2 \pi)}^{2} \mu \!\! \int \!\! \rd \up \, {| \psid_{\rm tot} |}^{2} \, \deltaD (\Omega (u) - \Omega (\up)) \, F (\up) ,
\label{def_D1_D2}
  \\
  D_{1} (u) & \!=\! \frac{1}{2} \frac{\partial D_{2}}{\partial u} + 2 \pi^{2} \mu \!\! \int \!\! \rd \up \, {| \psid_{\rm tot} |}^{2} \, \deltaD (\Omega (u) - \Omega (\up)) \, \frac{\partial F}{\partial \up} .
              \nonumber
\end{align}
In practice, for a given value of $u$, one can carry out the integral ${ \! \int \! \rd \up }$ in Eq.~\eqref{BL_Eq} by finding the resonant actions $u_{*}$ satisfying ${ \Omega (u_{*}) = \Omega (u) }$, which allows for the replacement ${ \deltaD (\Omega (u) - \Omega (u_{*})) = \sum_{u_{*}} \!\! \deltaD (u \!-\! u_{*})/|\partial \Omega / \partial u|_{u = u_{*}} }$

Because of its prefactor ${ \mu = \Mtot/N }$, the \BL\ equation describes the
system's long-term self-consistent evolution computed at first-order in the
${1/N}$ effects, accounting for the amplification by collective effects.  Here,
it is important to note that (i) the orbital space is one dimensional (cf.\ the
one dimensional integral ${ \! \int \! \rd \up }$ in Eq.~\eqref{BL_Eq}), (ii)
the symmetry of the interaction imposes ${1\!:\!1}$ resonances (cf.\ the
absence of sums over ${(k , \kp)}$ in Eq.~\eqref{BL_Eq}). As a consequence, if
the system's mean frequency profile, ${ u \mapsto \Omega (u) }$, is monotonic,
the resonance condition ${ \deltaD (\Omega (u) - \Omega (\up)) }$ only allows
for local resonances, i.e.\ ${ \up = u }$, leading to zero flux and
${ \partial F / \partial t = 0 }$.  In that case, the system cannot relax under
${1/N}$
effects~\citep{BarreGupta2014,Chavanis2013EPJP,Rocha2014,Lourenco2015}. It
undergoes a so-called kinetic blocking~\citep{Chavanis2007}, and can only relax
under weaker finite-$N$ effects associated with higher-order correlations.
Still, even if the flux vanishes, the diffusion coefficients ${ D_{1} (u) }$
and ${ D_{2} (u) }$ remain non-zero. Conversely, for a non-monotonic frequency
profile, non-local resonances, ${ \up \neq u }$, are allowed, the flux is
non-zero, and the system can relax at the order ${1/N}$.  We illustrate these
various effects in the coming section.  Finally, we emphasize that the
Boltzmann \DF\, ${ F \propto \re^{- \beta H_{0} (u)} }$ is always a stationary
solution of the \BL\ equation. Yet, the fact that for kinetically blocked systems
any axisymmetric \DF\ is a stationary solution of the \BL\ equation does not imply
that these states remain stationary when higher order correlation effects are accounted for.

\section{Application}
\label{sec:Application}

We now illustrate the previous formalism, and compare it with direct $N$-body
simulations (whose details are presented in Appendix~\ref{sec:Nbody}).

\begin{figure}
\begin{center}
\includegraphics[width=0.48\textwidth]{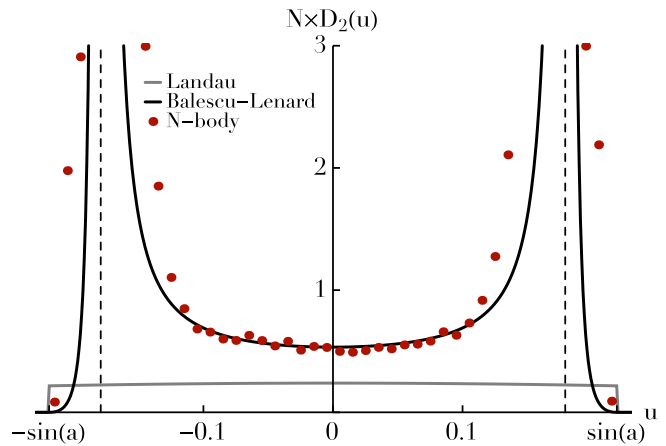}
\caption{Illustration of the second-order diffusion coefficient, ${ N \!\times\! D_{2} (u) }$, for the waterbag \DF\ from Eq.~\eqref{waterbag_DF}, as predicted by Eq.~\eqref{def_D1_D2}, in the absence (Landau) or presence (\BL\@) of collective effects, and compared with $N$-body simulations (that naturally include collective effects). As a result of the presence of neutral modes (see Eq.~\eqref{exp_D2_waterbag}), the \BL\ diffusion coefficient locally diverge, as indicated by the vertical dashed lines. 
  \label{fig:DWaterbag}}
\end{center}
\end{figure}

Following~\cite{GuptaMukamel2011}, we first consider a system
driven by interactions of the form
\begin{equation}
U (x) = - \alpha_{1} P_{1} (x) ; \;\;\; \Uext (u) = \Dext u^{2} ,
\label{monotonic_case}
\end{equation}
with ${ \alpha_{1} = 1 }$, ${ \Dext = 15 }$, and ${ P_{1} (x) = x }$. In that case,
Eq.~\eqref{mean_Hamiltonian} gives that ${ \Omega (u) }$ is a first degree
polynomial in $u$, i.e.\ the frequency profile is monotonic.  As for the
system's \DF\@, we consider a waterbag \DF\
\begin{equation}
F (u) = C \, \Theta \big( \sin (a) - |u| \big) ,
\label{waterbag_DF}
\end{equation}
with ${ \Theta (x) }$ the Heaviside function, and $C$ a normalisation constant.
We pick ${ \epsilon = \Dext \sin^{2} (a) / 3 = 0.24 }$, for which the system is
linearly stable~\citep{GuptaMukamel2011}.
The gradient of this \DF\ involves
Dirac deltas, which makes the computation of the response matrix immediate, as
one can get rid of the integral from Eq.~\eqref{Fourier_M},
and we refer to Eq.~\eqref{epsilon_Heisenberg} for the associated explicit expression.
Yet, because of these infinite gradients, the system also supports neutral modes (i.e.\ modes with
zero growth rates~\citep{Chavanis2005}), which lead to localized divergences in
the system's diffusion coefficients, as detailed in Eq.~\eqref{exp_D2_waterbag}.
In Fig.~\ref{fig:DWaterbag}, we illustrate
the \BL\ prediction for such diverging diffusion coefficients as well as
measurements from direct $N$-body simulations (using the procedure described in
Appendix~\ref{sec:Nbody}).

\begin{figure}
\begin{center}
\includegraphics[width=0.48\textwidth]{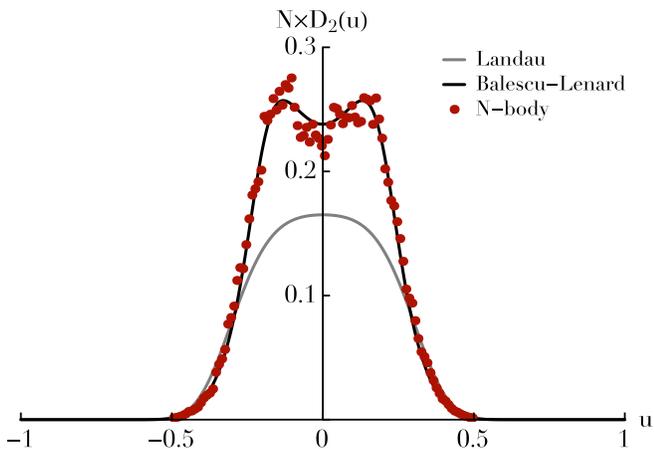}
\caption{Same as in Fig.~\ref{fig:DWaterbag}, but for the \DF\ from Eq.~\eqref{monotonic_DF}.
\label{fig:DMonotonic}}
\end{center}
\end{figure}

Keeping the same interactions as in Eq.~\eqref{monotonic_case}, one can avoid
the presence of neutral modes by considering a smooth \DF\@, for example
\begin{equation}
F (u) = C \, \re^{- (u/\sigma)^{4}} ,
\label{monotonic_DF}
\end{equation}
with ${ \sigma = 0.35 }$ and $C$ a normalisation constant.
In Appendix~\ref{sec:CompMat}, we present our implementation of the matrix method, and check that such a system is linearly stable (see Fig.~\ref{fig:NyquistContoursMonotonic}).
In Fig.~\ref{fig:DMonotonic}, we illustrate the \BL\ diffusion coefficients and the associated $N$-body measurements
for that system.

\begin{figure}
\begin{center}
\includegraphics[width=0.48\textwidth]{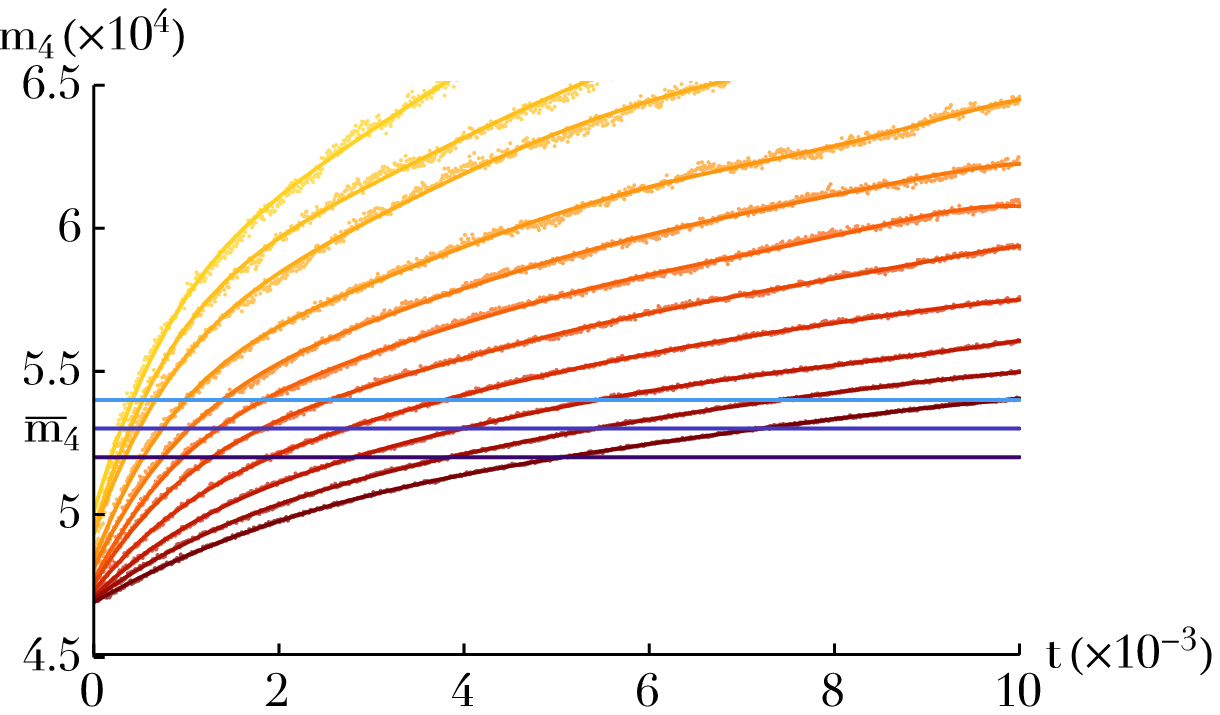}
\includegraphics[width=0.48\textwidth]{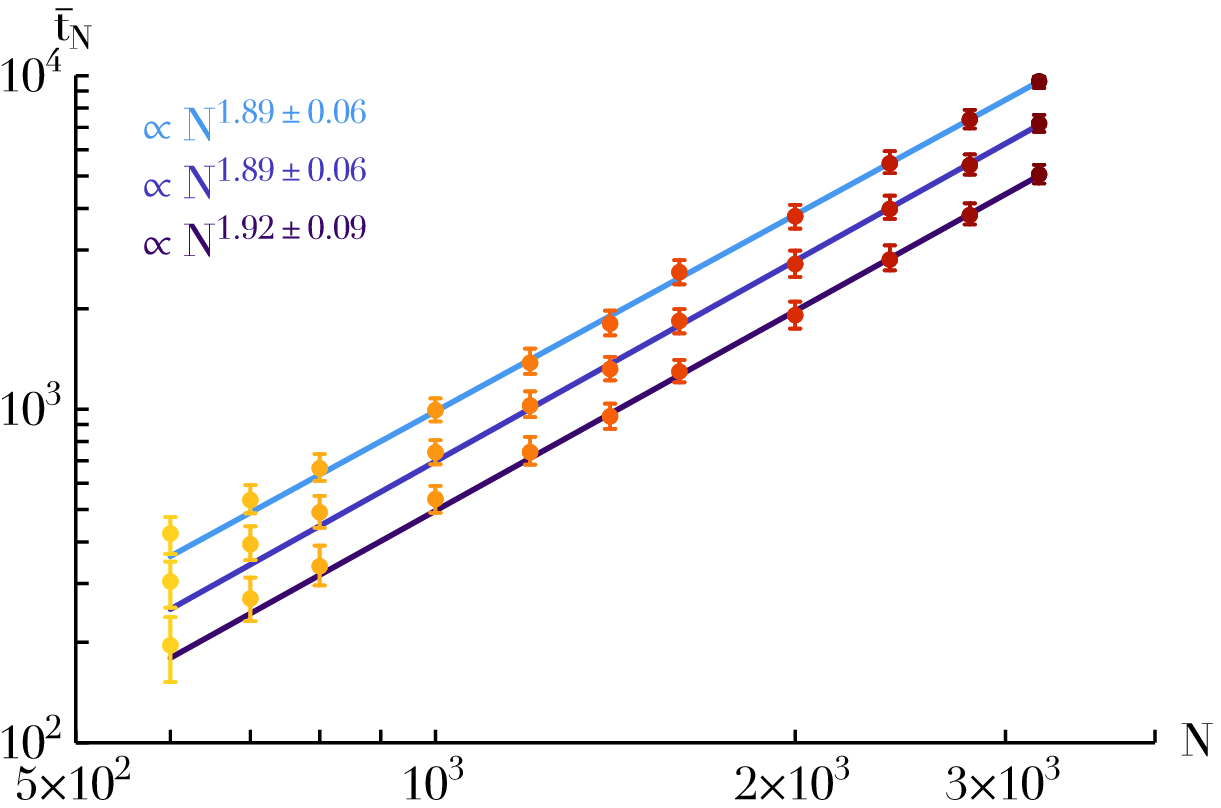}
\caption{Dependence of the relaxation rate with the number of particles for the waterbag system from Eq.~\eqref{waterbag_DF} that has a monotonic frequency profile yielding a kinetic blocking.
\textit{Top panel}: Time dependence of ${ m_{4} (N , t) }$ for simulations with ${ N \!\in\! \{ 6, 7, 8, 10, 12, 14, 16, 20, 24, 28, 32 \} \times 10^{2} }$ (from light to dark colors) averaged over $100$ realisations (dots), and the associated fits (curves).
The horizontal lines represent the threshold values $\om_{4}$ used to measure the crossing times.
\textit{Bottom panel}: Dependence of the crossing time $\ot_{N}$ with the number of particles for different $\om_{4}$ (light to dark colors). Errors bars for the crossing times were estimated by performing ${200}$ bootstrap resamplings over the realisations available: colored dots represent the median value, and error bars the ${10\%}$ and ${90\%}$ confidence levels. Errors on the power-law fits were estimated by fitting each bootstrap resamplings with a power-law, while the plotted fits are the best fit for the median values.
\label{fig:ScalingWaterbag}}
\end{center}
\end{figure}

Glancing at Eq.~\eqref{BL_Eq}, we argued that a system with a monotonic frequency profile undergoes a kinetic blocking and cannot relax under ${1/N}$ effects. We illustrate this in Fig.~\ref{fig:ScalingWaterbag} for the waterbag \DF\ from Eq.~\eqref{waterbag_DF}.
Following~\cite{Lourenco2015}, the dependence of the relaxation rate with $N$ is estimated through the quantity
${ m_{4} (N , t) = \{ (u - \{ u \})^{4} \} }$, with ${ \{ x \} = \sum_{i} x_{i} / N }$ the average over all the particles of a given realisation.
For a given $N$, the time series of ${ m_{4} (N , t) }$ is averaged over ${100}$ realisations, as illustrated in the top panel of Fig.~\ref{fig:ScalingWaterbag}. Finally, for a given threshold value $\om_{4}$, we determine the crossing time ${ \ot_{N} }$ such that ${ m_{4} (N , \ot_{N}) = \om_{4} }$. Should the \BL\ equation~\eqref{BL_Eq} have a non-vanishing flux, one expects the scaling ${ \ot_{N} \propto N }$. The dependence of ${ N \mapsto \ot_{N} }$ for the waterbag \DF\ is illustrated in the bottom panel of Fig.~\ref{fig:ScalingWaterbag}. In the range ${ 6 \!\times\! 10^{2} \leq N \leq 32 \!\times\! 10^{2} }$, we measure the scaling ${ \ot_{N} \propto N^{1.92 \pm 0.09} }$, which is expected to converge to $N^{2}$ for larger values of $N$~\citep{Lourenco2015}. This system indeed suffers from a kinetic blocking because of the impossibility of non-local resonant couplings for a monotonic frequency profile.

\begin{figure}
\begin{center}
\includegraphics[width=0.48\textwidth]{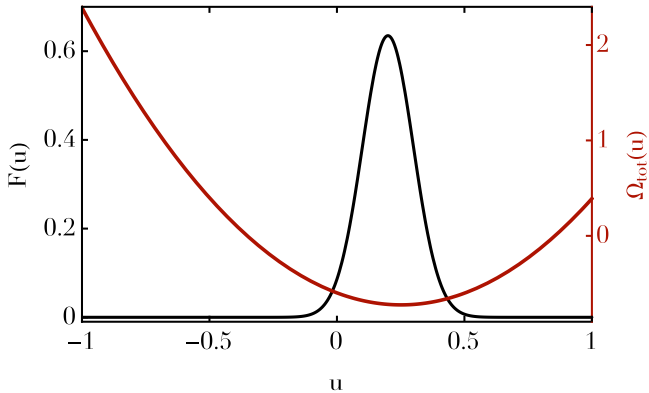}
\caption{Illustration of the \DF\ from Eq.~\eqref{nonmonotonic_DF} and the associated non-monotonic frequency profile ${ u \mapsto \Omega (u) }$. In the region of the \DF\@'s maximum, the resonance condition ${ \Omega (\up) = \Omega (u) }$ has two solutions, allowing for non-local resonant couplings.
\label{fig:DFOmegaNonMonotonic}}
\end{center}
\end{figure}

\begin{figure}
\begin{center}
\includegraphics[width=0.48\textwidth]{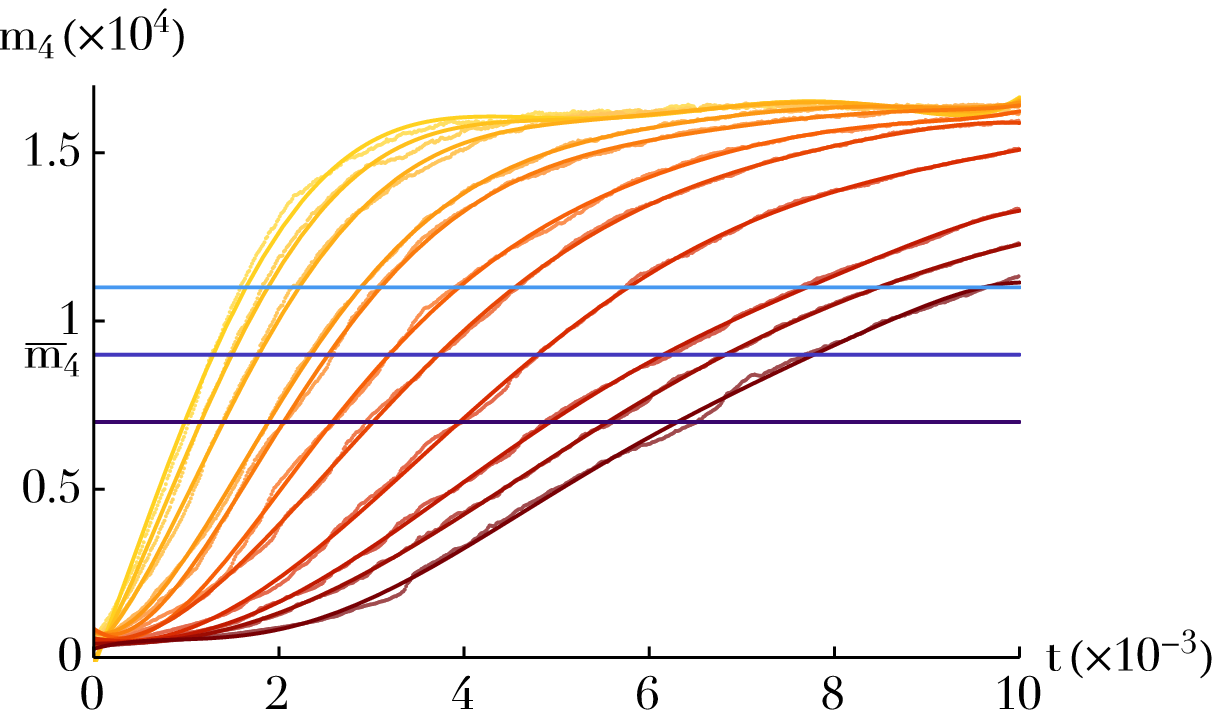}
\includegraphics[width=0.48\textwidth]{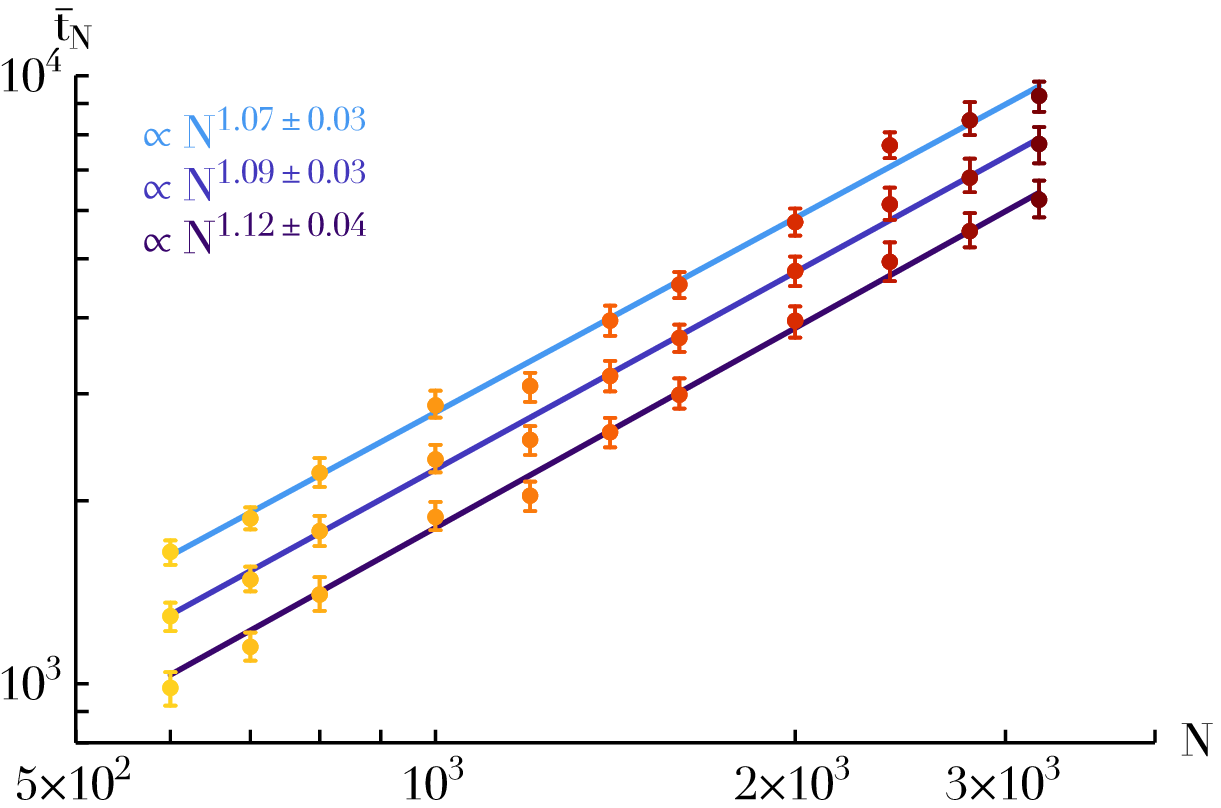}
\caption{Same as in Fig.~\ref{fig:ScalingWaterbag}, but for the \DF\ from Eq.~\eqref{nonmonotonic_DF} that has a non-monotonic frequency profile, preventing any kinetic blocking.
\label{fig:ScalingNonMonotonic}}
\end{center}
\end{figure}

In order to recover the scaling predicted by the \BL\ equation (while assuming that ${ \Uext (u) = \Dext u^{2} }$ as in Eq.~\eqref{monotonic_case}), one has to consider a model in which higher harmonics (${ \ell = 3 }$ or higher)
contribute to the pairwise interaction.
To illustrate this point, we finally consider a system driven by interactions of the form
\begin{equation}
U (x) = - \alpha_{1} P_{1} (x) - \alpha_{3} P_{3} (x) ; \;\;\; \Uext (u) = \Dext u^{2} ,
\label{nonmonotonic_case}
\end{equation}
with ${ \alpha_{1} \!=\! \alpha_{3} \!=\! 1 }$, ${ \Dext \!=\! - 1/2 }$,
and ${ P_{3} (x) \!=\! \tfrac{1}{2} (5x^{3} \!-\! 3x) }$.
In that case, Eq.~\eqref{mean_Hamiltonian} gives that ${ \Omega (u) }$
is a non-monotonic second degree polynomial in $u$.
We choose the system's axisymmetric \DF\ to be
\begin{equation}
F (u) = C \, \re^{- (u - u_{0})^{2} / (2 \sigma^{2})} ,
\label{nonmonotonic_DF}
\end{equation}
with ${ u_{0} = 0.2 }$ and ${ \sigma = 0.1 }$, and illustrate it in Fig.~\ref{fig:DFOmegaNonMonotonic}.
Following Appendix~\ref{sec:CompMat}, we checked that such a system is linearly stable.
In Fig.~\ref{fig:ScalingNonMonotonic}, we estimate the scaling of the system's relaxation with the number of particles.
In the range ${ 6 \!\times\! 10^{2} \leq N \leq 32 \!\times\! 10^{2} }$, we measure a scaling of the form ${ \ot_{N} \propto N^{1.1 \pm 0.06} }$, that is in sensible agreement with the prediction from the \BL\ equation. Because this system can support non-local orbital resonances, it relaxes much more efficiently than kinetically-blocked systems.

\section{Conclusion}
\label{sec:Conclusion}

The inhomogeneous \BL\ equation is being increasingly used to constrain complex dynamical regimes, such as the ${1D}$ \HMF\ model~\citep{Benetti2017}, ${2D}$ razor-thin stellar disks~\citep{Fouvry2015}, or ${ 3D }$ stellar systems with or without central mass~\citep{BarOr2018,Hamilton2018}.

In the present letter, we illustrated how the same method may be applied to characterize the dynamics of long-range coupled particles on a sphere in the axisymmetric limit. Once one has recognized that this system's evolution equations are formally identical to the ones of a long-range interacting integrable system, the derivation of the kinetic theory becomes straightforward.
In the present case, the reduced number of dimensions of phase space imposes additional geometrical constraints to the system's dynamics, e.g.\ allowing only for ${1\!:\!1}$ resonance. We detailed how in the presence of a monotonic frequency profile, the system is submitted to a kinetic blocking and cannot relax under ${1/N}$ effects, a behavior already encountered in the context of axisymmetrically distributed point vortices~\citep{Chavanis2007}.\footnote{
Because of the absence of a (quadratic) kinetic energy term in the Hamiltonian from Eq.~\eqref{def_Hamiltonian}, the present model shares some similarities with ${2D}$ point vortices, e.g.\ the existence of negative temperature statistical equilibria, or a similar \BL\ equation for axisymmetric distributions of point vortices.
Depending on the harmonic indices present in the interaction potential, the shape of ${ \Omega (u) }$ for the present model can be independent of time, while in the vortex case it depends on time as it is obtained self-consistently from the system's density profile~\citep{Chavanis2007}.
In particular, point vortices systems can still undergo a kinetic blocking even if the frequency profile is initially non-monotonic,
provided that it becomes monotonic during the evolution.
}
This blocking gets lifted in the presence of a non-monotonic frequency profile, for which non-local resonant couplings are possible.

To emphasize the strength of the \BL\ formalism, we presented quantitative comparisons with direct numerical simulations, recovering both the individual diffusion coefficients, as well as the expected scaling of the relaxation rate with the number of particles.

Despite its recent success, the kinetic theory of long-range interacting systems still asks for more developments, in particular to describe systems with fully degenerate frequency profiles (i.e.\ ${ \Omega (\bJ) = 0 }$, e.g.\ in the isotropic limit of the present system), or to obtain the ${1/N^{2}}$ kinetic equation for systems undergoing a kinetic blocking~\citep{Rocha2014,Lourenco2015}.

\acknowledgments

JBF acknowledges support from Program number HST-HF2-51374 which was provided
by NASA through a grant from the Space Telescope Science Institute, which is
operated by the Association of Universities for Research in Astronomy,
Incorporated, under NASA contract NAS5--26555.  BB is supported by membership
from Martin A. and Helen Chooljian at the Institute for Advanced Study.

\appendix

\section{The inhomogeneous Balescu--Lenard equation}
\label{sec:GenericBL}

In this Appendix, we repeat the main results regarding the inhomogeneous \BL\ equation, first derived in~\cite{Heyvaerts2010,Chavanis2012}.
These results are used in the main text to concisely derive the kinetic equation for the problem at hand.

We generically consider an Hamiltonian system in $2d$ dimensions, and write the
phase space canonical coordinates as ${ \bw = (\bT , \bJ) }$, respectively the angle and action coordinates~\citep{BinneyTremaine2008}.
The system is assumed to be in an integrable steady state, and following Jeans' theorem~\citep{Jeans1915}, it can be described by a \DF\ of the form, ${ F (\bw) = F (\bJ) }$, that we normalize as ${ \! \int \! \rd \bw \, F = M_{\rm tot} }$. 
We denote the mean (integrable) potential as ${ H_{0} (\bw) = H_{0} (\bJ) }$, with the associated orbital frequencies ${ \bO (\bJ) = \partial H_{0} / \partial \bJ }$.
The system comprises $N$ particles of mass ${ \mu = M_{\rm tot} / N }$, coupled one to another through a long-range interaction potential ${ U (\bw , \bwp) }$.

As a result of the finite number of particles, the system's orbital structure
gets slowly distorted. To first order in ${1/N}$  this dynamics is described by the inhomogeneous \BL\ equation
\begin{align}
& \frac{\partial F (\bJ)}{\partial t} = \pi {(2 \pi)}^{d} \mu \,
  \frac{\partial }{\partial \bJ} \cdot \bigg[ \sum_{\bk , \bkp} \bk \!\! \int \!\! \rd \bJp \, {\big| \psid_{\bk\bkp} (\bJ , \bJp , \bk \cdot \bO (\bJ)) \big|}^{2}
\nonumber
\\
& \times \deltaD (\bk \cdot \bO (\bJ) \!-\! \bkp \cdot \bO (\bJp)) \bigg(\! \bk \cdot \frac{\partial }{\partial \bJ} \!-\! \bkp \cdot \frac{\partial }{\partial \bJp} \!\bigg) F (\bJ) F (\bJp) \bigg].
\label{generic_BL}
\end{align}

Following Kalnajs matrix method~\citep{Kalnajs1976}, one introduces a biorthogonal basis of potentials and densities ${ (\psi^{(p)} (\bw) , \rho^{p} (\bw)) }$, with the convention
\begin{align}
& \, \psi^{(p)} (\bw) = \!\! \int \!\! \rd \bwp \, U (\bw , \bwp) \, \rho^{(p)} (\bwp) ,
\nonumber
\\
& \, \!\! \int \!\! \rd \bw \, \rho^{(p)} (\bw) \, \psi^{(q) *} (\bw) = - \delta_{p}^{q} .
\label{biorthogonal_condition}
\end{align}
The pairwise interaction, ${ U (\bw , \bwp) }$, can then be cast under the separable form
\begin{equation}
U (\bw , \bwp) = - \sum_{p} \psi^{(p)} (\bw) \, \psi^{(p) *} (\bwp) .
\label{separable_U}
\end{equation}
The separable decomposition from Eq.~\eqref{separable_U} is a fundamental equation that determines what are the \textit{natural} basis elements appropriate for a given problem, e.g.\ as in Eq.~\eqref{def_psip}.
The bare susceptibility coefficients, ${ \psi_{\bk\bkp} (\bJ , \bJp) }$, are the Fourier transform of the pairwise interaction ${ U (\bw , \bwp) }$ w.r.t.\ ${ (\bT , \bTp) }$, namely
\begin{align}
U (\bw , \bwp) & \, =  \sum_{p} \sum_{\bk , \bkp \in \mathbb{Z}^{d}} \psi_{\bk\bkp} (\bJ , \bJp) \, \re^{\ri (\bk \cdot \bT - \bkp \cdot \bTp)} ,
\label{Fourier_U}
\\
\psi_{\bk\bkp} (\bJ , \bJp) & \, = \!\! \int \!\! \frac{\rd \bT}{(2 \pi)^{d}} \frac{\rd \bTp}{(2 \pi)^{d}} \, U (\bT , \bJ , \bTp , \bJp) \, \re^{- \ri (\bk \cdot \bT - \bkp \cdot \bTp)} ,
\nonumber
\end{align}
and Eq.~\eqref{generic_BL} with ${ \psid_{\bk\bkp} (\bJ , \bJp , \omega) \to \psi_{\bk\bkp} (\bJ , \bJp) }$ is the inhomogeneous Landau equation.
One can write
\begin{equation}
\psi_{\bk\bkp} (\bJ , \bJp) = - \sum_{p} \psi^{(p)}_{\bk} (\bJ) \, \psi_{\bkp}^{(p) *} (\bJp) ,
\label{bare_psi}
\end{equation}
with ${ \psi_{\bk}^{(p)} (\bJ) \!=\! \! \int \! \rd \bT / (2 \pi)^{d} \psi^{(p)} (\bw) \, \re^{- \ri \bk \cdot \bT} }$ standing for the Fourier transform of the basis elements.
To account for collective effects (i.e.\ the spontaneous amplification of fluctuations),
one replaces the bare susceptibility coefficients by their dressed analogs
\begin{equation}
\psid_{\bk \bkp} (\bJ , \bJp , \omega) = - \sum_{p , q} \psi^{(p)}_{\bk} (\bJ) \, \big[ \bI - \wbM (\omega) \big]_{pq}^{-1} \, \psi^{(q) *}_{\bkp} (\bJp) ,
\label{dressed_psi}
\end{equation}
as introduced in the \BL\ Eq.~\eqref{generic_BL}.
Finally, in Eq.~\eqref{dressed_psi}, the response matrix of a long-range integrable system is given by
\begin{align}
\wbM_{pq} (\omega) = (2 \pi)^{d} \sum_{\bk} \!\! \int \!\! \rd \bJ \, \frac{\bk \cdot \partial F / \partial \bJ}{\omega - \bk \cdot \bO (\bJ)} \, \psi^{(p) *}_{\bk} (\bJ) \, \psi^{(q)}_{\bkp} (\bJp) ,
\label{Fourier_M_generic}
\end{align}
and its effective numerical computation is briefly illustrated in Appendix~\ref{sec:CompMat}.
We note that (i) Eq.~\eqref{bare_psi} can be obtained from
Eq.~\eqref{dressed_psi} by taking ${ \wbM = 0 }$ (i.e.\ switching off
collective effects); (ii) the substitution of Eq.~\eqref{separable_U} into
Eq.~\eqref{biorthogonal_condition} leads to an identity; (iii)
Eq.~\eqref{separable_U} can be obtained from Eqs.~\eqref{Fourier_U}
and~\eqref{bare_psi} showing the unicity of this expression.
As emphasized after Eq.~\eqref{Fourier_M}, the response matrix, ${ \wbM (\omega) }$,
characterizes the linear stability of the system.

\section{The case of the Heisenberg spins}
\label{sec:HeisenbergLimit}

In this Appendix, we consider the case where the system's interaction is limited to only one harmonic $\ell$, so that ${ U (x) = - \alpha_{\ell} P_{\ell} (x) }$.
This is of particular relevance for classical Heisenberg spins (${ \ell = 1 }$)~\citep{GuptaMukamel2011,BarreGupta2014}, and the Maier-Saupe model for liquid crystals (${ \ell = 2 }$)~\citep{MaierSaupe1958,RoupasKocsis2017}.
In that limit, the dressed susceptibility coefficients from Eq.~\eqref{suscp_coeff} become
\begin{equation}
\psi_{k \kp}^{\rd} (u , \up , \omega) = - \delta_{k}^{\kp} \, \frac{ c_{\ell}^{k} (u) \, c_{\ell}^{k *} (\up)}{ \veps_{\ell}^{k} (\omega)} ,
\label{suscp_coeff_one}
\end{equation}
with ${0 < |k| \leq \ell }$, and the susceptibility coefficient, ${ \veps_{\ell}^{k} (\omega) }$, follows from Eq.~\eqref{Fourier_M} reading
\begin{equation}
\veps_{\ell}^{k} (\omega) = 1 - 2 \pi \!\! \int \!\! \rd u \, \frac{k \, \partial F / \partial u}{\omega - k \Omega (u)} \, | c_{\ell}^{k} (u) |^{2} .
\label{def_veps}
\end{equation}

Let us now follow Eq.~\eqref{monotonic_case} and restrict ourselves to the pairwise interaction
${ U (x) = - \alpha_{1} P_{1} (x) }$, with ${ \alpha_{1} = 1 }$.
In that context, Eq.~\eqref{def_veps} reduces to
\begin{equation}
\veps_{1}^{\pm 1} (\omega) = 1 \mp \pi \!\! \int \!\! \rd u \, \frac{(1 - u^{2}) \partial F / \partial u}{\omega \mp \Omega (u)} ,
\label{epsilon_Heisenberg}
\end{equation}
with ${ \Omega (u) = h_{1} + 2 D_{\rm ext} u }$ (see Eq.~\eqref{mean_Hamiltonian} for the definition of $h_{\ell}$).
As such, in Eq.~\eqref{epsilon_Heisenberg}, we immediately recover the susceptibility coefficients obtained
by a more complex method in~\cite{BarreGupta2014} (see Eq.~{(33)} therein).
For a waterbag \DF\ as in Eq.~\eqref{waterbag_DF}, the expression of the susceptibility coefficients can be further simplified to become
\begin{equation}
\veps_{1}^{\pm 1} (\omega) = 1 + \frac{D_{\rm ext} \, (1 - \sin^{2} (a))}{\omega^{2} - {(2 D_{\rm ext} \sin (a))}^{2}} \equiv \veps (\omega) .
\label{epsilon_Heisenberg_waterbag}
\end{equation}
Such a system is linearly stable if there exists no ${ \omega = \omega_{0} + \ri \eta }$ (with ${ \eta > 0 }$) for which ${ \veps (\omega) = 0 }$. The constraint ${ \text{Im} [\veps (\omega)] = 0 }$, naturally imposes ${ \omega_{0} = 0 }$. As for the constraint ${ \text{Re} [\veps (\ri \eta) ] = 0 }$, and introducing the energy of the system as ${ \epsilon = D_{\rm ext} \sin^{2} (a)/3 }$, one concludes that the system admits no unstable modes if
\begin{equation}
\kappa = \frac{D_{\rm ext} - 3 \epsilon}{12 D_{\rm ext} \epsilon}
\label{def_kappa}
\end{equation}
satisfies ${ \kappa <1 }$.
Introducing the critical energy ${ \epsilon_{\star} \!=\! D_{\rm ext}/(3 \!+\! 12 D_{\rm ext}) }$, the system is therefore stable if ${ \epsilon > \epsilon_{\star} }$, and we recover the criterion put forward in~\cite{GuptaMukamel2011,BarreGupta2014}.

For a waterbag \DF\ from Eq.~\eqref{waterbag_DF}, one can finally compute explicitly the diffusion coefficient from Eq.~\eqref{def_D1_D2}. The frequency, ${ \Omega (u) = 2 D_{\rm ext} u }$, being monotonic, the resonance condition is straighforwardly solved, and Eq.~\eqref{waterbag_DF} gives
\begin{equation}
D_{2} (u) = \frac{{(2 \pi)}^{2} \mu}{D_{\rm ext}} \, \frac{{| c_{1} (u) |}^{4}}{{|\veps (\Omega (u))|}^{2}} \, F (u) ,
\label{exp_D2_waterbag}
\end{equation}
with ${ c_{1} (u) \!=\! \sqrt{(1-u^{2})/2} }$. Introducing ${ \tw \!=\! \omega / \omega_{\max} }$, with ${ \omega_{\max} \!=\! (2 D_{\rm ext} \sin (a)) }$, the inverse of the susceptibility coefficient reads
\begin{equation}
\frac{1}{\veps(\omega)} = \frac{1 - \tw^{2}}{(1 \!-\! \kappa) - \tw^{2}} ,
\label{inverse_suscept}
\end{equation}
where we recall that ${ \kappa < 1 }$ for a linearly stable system.
As a consequence, for ${ \tw = \sqrt{1 \!-\! \kappa} }$, the inverse of the susceptibility coefficient becomes infinite,
i.e.\ the system supports a neutral mode~\citep{Chavanis2005}.
This leads in particular to a divergence of the diffusion coefficient in ${ u \!=\! \pm \sin (a) \sqrt{1 \!-\! \kappa} }$, as highlighted in Fig.~\ref{fig:DWaterbag}.
We finally note that in the absence of collective effects (i.e.\ in the Landau limit), these divergences vanish.

\section{The $N$-body implementation}
\label{sec:Nbody}

In this Appendix, we present our $N$-body implementation of the problem at
hand. Glancing back at the equations of motion Eq.~\eqref{EOM_short}, one notes that the velocity vector, ${ \rd \bL_{i} / \rd t }$, is expressed only as a function of the current particle's location, ${ \bL_{i} (t) }$, and the instantaneous values of the magnetisations, ${ M_{\ell}^{m} (t) }$. There are $N$ such evolution equations, but since magnetisations are shared by all particles, their computation can be done only once per timestep. As a result, the overall complexity of advancing the particles for one timestep scales like ${ \mO (N \, \ell_{\max}^{2}) }$, with $\ell_{\max}$ the maximum harmonic number appearing in the considered pairwise interaction in Eq.~\eqref{Legendre_U}.

The heart of the $N$-body implementation is then (i) to compute efficiently the spherical harmonics (and the vector ones) at the location of the particles, (ii) to compute the magnetisations in Eq.~\eqref{def_Mlm}, and (iii) to compute the velocity fields in Eq.~\eqref{EOM_short}.
To compute the spherical harmonics, it is convenient to work with real spherical harmonics. These are computed following~\cite[][see Eq.~{(6.7.9)}]{Press2007} for the renormalized associated Legendre polynomials, and the second-order recurrence relation ${ \cos (m \phi) = 2 \cos (\phi) \cos ((m-1)\phi) - \cos ((m-2)\phi) }$ (similarly for ${ \sin (m \phi) }$) for the azimuthal component.
For the (real) vector spherical harmonics, we follow the recurrences presented in~\cite[][see Appendix~{(B.2)}]{Mignard2012}, adapted to the renormalized associated Legendre polynomials.
Once all velocity vectors ${ \rd \bL_{i} / \rd t }$ are determined, particles are advanced for a timestep $h$ (${ =10^{-3} }$ in all our applications) using a fourth-order Runge-Kutta integrator~\citep[][see Eq.~{(17.1.3)}]{Press2007}.

To measure diffusion coefficients in $N$-body simulations, we proceed similarly to~\cite{Benetti2017} in the case of the \HMF\ model.
We note that the present model shares some similarities with the \HMF\ model~\citep{AntoniRuffo1995}, in particular the property of being a decoupled $N$-body problem, i.e.\ it can be integrated in ${ \mO (N) }$ operations per timestep, rather than ${ \mO (N^{2} / 2) }$.
Here, to measure diffusion coefficients, we perform ${ N_{\rm real} =  200 }$ different realisations, with ${ N = 10^{5} }$ particles. At the initial time, particles are divided among action bins of size ${ \delta u = 10^{-2} }$. For each realisation and action bin, we determine the time series of the mean square variation of ${ \Delta u^{2} (t) = {(u(t) - u (0))}^{2} }$, averaged over all the particles initially within the bin. For every action bin, these time series are then averaged over all realisations, and considered up to the time where ${ \langle \Delta u^{2} (t) \rangle \geq {(\delta u)}^{2} }$. The diffusion coefficients are obtained finally through a linear fit of these ensemble-averaged series, as in Figs.~\ref{fig:DWaterbag} and~\ref{fig:DMonotonic}.

\section{The matrix method}
\label{sec:CompMat}

\begin{figure}
\begin{center}
\includegraphics[width=0.48\textwidth]{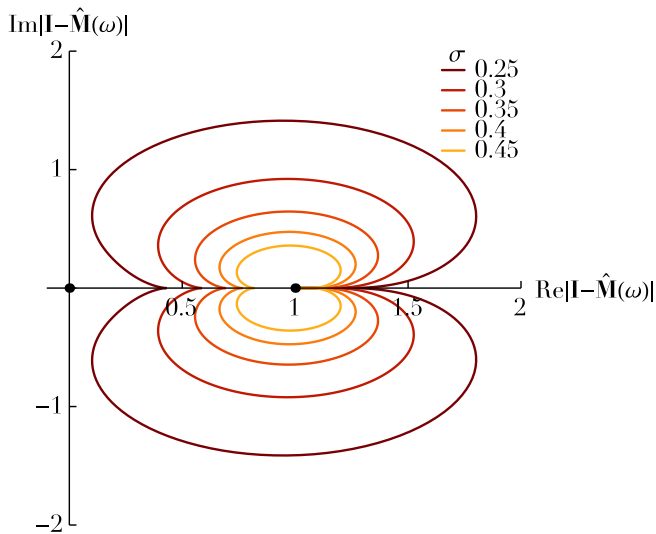}
\caption{Illustration of the Nyquist contours ${ \omega \mapsto \det [\bI - \wbM (\omega , \eta = 10^{-8})] }$ for the \DF\ from Eq.~\eqref{monotonic_DF} with different dispersions $\sigma$. None of these contours enclose the origin, indicating that these systems are linearly stable. The smaller the dispersion, the closer is the system to being unstable.
\label{fig:NyquistContoursMonotonic}}
\end{center}
\end{figure}

The generic expression of the response matrix is given by Eq.~\eqref{Fourier_M}. It asks to compute an expression of the form
\begin{align}
\!\! \int_{-1}^{1} \!\! \rd u \, \frac{g (u)}{h (u) + \ri \eta} & \, \simeq \sum_{i} \!\! \int_{- \frac{\delta u}{2}}^{\frac{\delta u}{2}} \!\!\!\! \rd x \, \frac{a_{g}^{i} + b_{g}^{i} x}{a_{h}^{i} + b_{h}^{i} x + \ri \eta}
\nonumber
\\
& \, = \sum_{i} \frac{a_{g}^{i}}{a_{h}^{i}} \, \delta u \, \alephD \big[ \tfrac{b_{g}^{i} \delta u}{a_{g}^{i}} , \tfrac{b_{h}^{i} \delta u}{a_{h}^{i}} , \tfrac{\eta}{a_{h}^{i}} \big] ,
\label{generic_Fourier_Mat}
\end{align}
where ${ g(u), h(u) }$ are real, and ${ \eta > 0 }$ is an imaginary part added to the frequency $\omega$. To get the r.h.s.\@ of Eq.~\eqref{generic_Fourier_Mat}, we followed~\cite{Fouvry2015}, truncated the integration domain ${ u \in [-1 ; 1] }$ into $K$ regions, introducing ${ \delta u = 2 / K }$, so that the center of each region is ${ u_{i} = \!-\! 1 \!+\! \delta u(i \!-\! \tfrac{1}{2}) }$ with ${ 1 \leq i \leq K }$, and performed a linear expansion of the numerator and denominator, so that ${ a_{g}^{i} = g (u_{i}) }$ and ${ b_{g}^{i} = \rd g (u_{i}) / \rd u }$ (similarly for $h$). To get the second line, we assumed ${ a_{g}, a_{h} \neq 0 }$, and introduced the dimensionless function $\alephD$
\begin{equation}
\alephD [b , c , \eta] = \!\! \int_{- \frac{1}{2}}^{\frac{1}{2}} \!\!\!\! \rd x \, \frac{1 + b x}{1 + c x + \ri \eta} = G[\tfrac{1}{2}] - G[- \tfrac{1}{2}] ,
\label{calc_alephD}
\end{equation}
where for the primitive ${ G (x) }$, one can choose
\begin{align}
G [ x ] = & \, \frac{b x}{c} + \frac{b \eta + \ri (c - b)}{2 c^{2}}
\label{expression_G}
\\
& \, \times \big\{ 2 \big( \tfrac{\pi}{2} - \tan^{-1} \big[ \tfrac{1 + c x}{\eta} \big] \big) - \ri \ln \big[ {(1 + c x)}^{2} + \eta^{2} \big] \big\} .
\nonumber
\end{align}
We illustrate this method in Fig.~\ref{fig:NyquistContoursMonotonic}, by representing the Nyquist contours associated with the system from Eq.~\eqref{monotonic_DF}. This shows that this particular system is linearly stable.
Throughout the applications presented in the main text, we truncated the orbital space in ${ K = 10^{4} }$ elements, and added to the frequency a small imaginary part ${ \eta = 10^{-8} }$ to regularize the resonant denominator. We checked that these choices had no impact on our results.

\end{document}